%% file: paper-revised-ms-only.tex
\documentclass[3p,times]{elsarticle}
\usepackage{amssymb}
\usepackage{epsfig}
\usepackage{mathrsfs}
\usepackage{graphicx}
\usepackage{amssymb}
\usepackage{amsmath}
\usepackage{lscape}
\usepackage{color}
\usepackage{dcolumn}
\usepackage{threeparttable}
\usepackage{booktabs}
\usepackage[labelfont=bf,singlelinecheck=false,font=footnotesize]{caption} 
\usepackage{hyperref}
\biboptions{square,sort&compress} 
\newcommand{\beq}{\begin{equation}}
\newcommand{\eeq}{\end{equation}}
\newcommand{\AP}[2]{${#1}\:^{#2}\text{A}^{\prime}$}
\newcommand{\APP}[2]{${#1}\:^{#2}\text{A}^{\prime\prime}$}

\journal{Chemical Physics Letters}

\begin{document}
\begin{frontmatter}
\title{
Extension of the Active-Orbital-Based and Adaptive CC(\protect\mbox{$P$;$Q$})
Approaches to Excited Electronic States: Application to Potential Cuts of Water
}
\author[label1]{Karthik Gururangan}
\author[label1]{Jun Shen}
\author[label1,label2]{Piotr Piecuch\corref{cor1}}
\address[label1]{Department of Chemistry, Michigan State University, 
East Lansing, Michigan 48824, USA}
\address[label2]{Department of Physics and Astronomy, Michigan State University, East Lansing, Michigan 48824, USA}
\ead{piecuch@chemistry.msu.edu}
\cortext[cor1]{Corresponding author}

\begin{abstract}
We report the first study using active-orbital-based and adaptive CC($P$;$Q$) 
approaches to describe excited electronic states. These CC($P$;$Q$) methodologies 
are applied, alongside their completely renormalized (CR) coupled-cluster (CC) 
and equation-of-motion (EOM) CC counterparts, to recover the ground- and 
excited-state potential cuts of the water molecule along the O--H bond-breaking coordinate 
obtained in the parent CC/EOMCC calculations with a full treatment of singles, doubles,
and triples (CCSDT/EOMCCSDT). We demonstrate that the active-orbital-based and adaptive 
CC($P$;$Q$) approaches closely approximate the CCSDT/EOMCCSDT data using
significantly reduced computational costs while improving the CR-CC and CR-EOMCC 
energetics in stretched regions of the O--H bond-breaking potentials.
\end{abstract}



\begin{keyword}
Coupled-Cluster Theory
\sep Equation-of-Motion Coupled-Cluster Formalism
\sep Completely Renormalized Coupled-Cluster Approaches 
\sep Active-Space Coupled-Cluster Methods
\sep Active-orbital-based CC($P$;$Q$) Approaches
\sep Adaptive CC($P$;$Q$) Methodology
\sep Ground- and Excited-State Potential Surfaces of Water
\end{keyword}

\end{frontmatter}


%
\section{Introduction}
\label{sec1}
The development of accurate and computationally practical methods for 
describing excited electronic states and potential energy surfaces (PESs) 
of molecules is a vital component of quantum chemistry. This task becomes 
especially challenging when examining excited states dominated by two- or 
other many-electron transitions and larger regions of excited-state PESs,
which are relevant to spectroscopic and photochemical 
applications.
It is nowadays well established that the excited-state extensions of the 
coupled-cluster (CC) theory 
\cite{cizek1,cizek2,cizek4} 
belonging to the equation-of-motion (EOM)
\cite{emrich,eomcc3}, 
linear-response (LR)
\cite{monk,mukherjee_lrcc,lrcc4,jorgensen}, 
and symmetry-adapted-cluster configuration interaction (CI) 
\cite{sacci4}
hierarchies are capable of providing reliable and systematically 
improvable
description of excited electronic states using polynomial computational steps 
within a conceptually straightforward single-reference (SR) ansatz.

In the EOMCC
framework, which is a focus of the present study, the exact excited states of the
$N$-electron system are defined as
$|\Psi_{\mu}\rangle = R_{\mu} |\Psi_{0}\rangle$, where
$|\Psi_{0}\rangle = e^{T}|\Phi\rangle$ is the ground-state CC wave function,
with $|\Phi\rangle$ designating the reference determinant that serves as
a Fermi vacuum, and $T = \sum_{n=1}^{N} T_{n}$ and
$R_{\mu} = \sum_{n=0}^{N} R_{\mu,n} = r_{\mu,0}\textbf{1} + \sum_{n=1}^{N} R_{\mu,n}$
($\mu > 0$)
are the cluster and EOM excitation operators, with $T_{n}$ and $R_{\mu,n}$ 
representing their $n$-body components ({\bf 1} is the identity operator).
The EOMCC approximations obtained by truncating $T$ and $R_{\mu}$ at a
given many-body rank provide a transparent route for computing
increasingly accurate and systematically improvable excited-state energetics
and properties, but the leading and most practical EOMCC approach with
singles and doubles (EOMCCSD) \cite{eomcc3}, obtained by truncating
$T$ and $R_{\mu}$ at their two-body components,
cannot handle multireference (MR) correlation effects
characterizing excited states dominated by two or other many-electron transitions and 
excited-state potentials outside the Franck--Condon region, especially when chemical bonds are
significantly stretched or broken
\cite{eomccsdt1,eomccsdt2,water}.
In fact, EOMCCSD is not
fully quantitative in describing singly excited states either \cite{creomcc-2015}.
In general, higher-level EOMCC methodologies that incorporate $T_{n}$ and 
$R_{\mu,n}$ components with $n>2$, such as the EOMCC approach with a full
treatment of singles, doubles, and triples (EOMCCSDT) \cite{eomccsdt1,eomccsdt2,eomccsdt3},
in which $T$ and $R_{\mu}$ are truncated at $T_{3}$ and $R_{\mu,3}$, respectively,
are necessary to obtain an accurate and more robust description.

Inspired by Refs.\ \cite{nmapp3,water}, in this work we focus on converging the
ground- and excited-state PESs of the water molecule along the O--H
bond-breaking coordinate corresponding to the ${\rm H}_{2}{\rm O} \rightarrow {\rm H} + {\rm OH}$
dissociation obtained with EOMCCSDT and its ground-state CCSDT counterpart
\cite{ccfullt,ccfullt2}. As shown in this study, the CCSDT/EOMCCSDT
water potentials are nearly exact, closely matching the full CI data reported in
Refs.\ \cite{nmapp3,water}, but the application of the full CCSDT and EOMCCSDT
methods, especially when larger molecules and basis sets are considered, is
hindered by the expensive computational steps that scale as
$\mathscr{N}^8$ with system size $\mathscr{N}$. To address this concern,
a variety of approximate EOMCC and LRCC schemes aimed at reducing the
prohibitive computational costs of the EOMCCSDT and LRCCSDT calculations
have been developed. Among them are methods that rely on perturbative arguments
to correct the EOMCCSD or LRCCSD calculations for $T_{3}$ and $R_{\mu,3}$
effects in an iterative or noniterative fashion \cite{eomap3,ccsdr3a,eomap5,cc3_1},
but these kinds of approaches, while accurate for excited states dominated
by one-electron transitions, struggle when the more MR excited states dominated by
two-electron transitions are examined. The newer triples corrections to
EOMCCSD \cite{eomccpt,eomccpt2,kkppeom,crccl_molphys,crccl_ijqc2,water,7hq,creomcc-2015},
such as the left-eigenstate completely renormalized (CR) EOMCC approach abbreviated as
CR-EOMCC(2,3) \cite{crccl_molphys,crccl_ijqc2,jspp-chemphys2012} and its rigorously size-intensive
$\delta$-CR-EOMCC(2,3) modification \cite{7hq,creomcc-2015} that rely on the moment
energy expansions \cite{kkppeom,crccl_molphys,crccl_ijqc2,jspp-chemphys2012,leszcz,ren1,kkppeom2,%
crccl_jcp,crccl_cpl}, are more robust in this regard
\cite{water,creom-azulene,creomcc-2015}, but they may still encounter difficulties
when $T_{3}$ and $R_{\mu,3}$ correlations become larger, nonperturbative, and
strongly coupled to their lower-rank $T_{1}$, $T_{2}$, $R_{\mu,1}$, and $R_{\mu,2}$ 
counterparts
\cite{jspp-chemphys2012,jspp-jcp2012,stochastic-ccpq-molphys-2020}.
This becomes evident when examining stretched regions of certain
excited-state water potentials with CR-EOMCC(2,3) \cite{water} (cf.
Sec.\ \ref{sec3.3}).

A robust solution to the above deficiencies of the perturbative corrections to EOMCCSD
and their CR-EOMCC(2,3) and similar counterparts is offered by the CC($P$;$Q$) framework
\cite{jspp-chemphys2012,jspp-jcp2012,%
stochastic-ccpq-prl-2017,%
stochastic-ccpq-molphys-2020,%
cipsi-ccpq-2021,adaptiveccpq2023},
which generalizes the original CR-CC \cite{leszcz,ren1,crccl_jcp,crccl_cpl,crccl_molphys,crccl_jpc}
and CR-EOMCC \cite{kkppeom,crccl_molphys,crccl_ijqc2,7hq,creomcc-2015} approaches
to unconventional truncations in the $T$ and $R_{\mu}$ operators that incorporate
the leading contributions to the $T_{n}$ and $R_{\mu,n}$ components with $n>2$
into the iterative parts of the computations, correcting the resulting energies
for the missing correlation effects of interest using moment expansions
similar to those used in CR-CC/EOMCC. In particular, the CC($P$;$Q$) formalism
provides a flexible and powerful framework for converging high-level CC/EOMCC energetics,
such as CCSDT, CCSDTQ \cite{ccsdtq0,ccsdtq2}, EOMCCSDT, etc., at small fractions of the
computational costs. The existing CC($P$;$Q$) methods aimed at recovering the CCSDT
and EOMCCSDT energetics include the active-orbital-based CC(t;3) approach \cite{jspp-chemphys2012,%
jspp-jcp2012}, which corrects the CCSDt
\cite{semi0b,semi2,semi4,piecuch-qtp}
and EOMCCSDt \cite{eomccsdt1,eomccsdt2,piecuch-qtp} calculations for
those $T_{3}$ and $R_{\mu,3}$ correlations that are missing in CCSDt/EOMCCSDt,
and its more black-box semi-stochastic \cite{stochastic-ccpq-prl-2017,%
stochastic-ccpq-molphys-2020}, selected-CI-driven \cite{cipsi-ccpq-2021},
and adaptive \cite{adaptiveccpq2023} descendants. The latter approach, which, along with
CC(t;3), is of interest in this work, allows one to converge high-level CC/EOMCC energetics
using a recursively defined sequence of CC($P$;$Q$) calculations guided by the intrinsic
structure of the CC($P$;$Q$) moment expansions without any reference to active orbitals
or non-CC concepts. Although the CC($P$;$Q$) theory provides a general framework for
describing ground as well as excited electronic states, so far only the semi-stochastic variant of
it has been tested on excited-state energetics \cite{stochastic-ccpq-molphys-2020}.
The present study helps to address this situation by applying the active-orbital-based CC(t;3) and
adaptive CC($P$;$Q$) approaches targeting CCSDT and EOMCCSDT to the ground- and excited-state
PESs of water corresponding to the ${\rm H}_{2}{\rm O} \rightarrow {\rm H} + {\rm OH}$
dissociation.
\section{Theory}
\label{sec2}
Each CC($P$;$Q$) calculation requires defining two disjoint subspaces of the
$N$-electron Hilbert space, referred to as the $P$ space $\mathscr{H}^{(P)}$
and $Q$ space $\mathscr{H}^{(Q)}$. The former space is spanned
by the excited determinants $|\Phi_K\rangle = E_K|\Phi\rangle$, with $E_K$ 
designating the elementary particle--hole excitation operator generating
$|\Phi_K\rangle$ from the reference determinant $|\Phi\rangle$,
which together with $|\Phi\rangle$
dominate the ground and excited electronic states of interest. The 
complementary subspace $\mathscr{H}^{(Q)}$ contains the excited determinants used 
to construct noniterative corrections to the energies
obtained in the CC and EOMCC calculations in the $P$ space, abbreviated as 
CC($P$) for the ground state and EOMCC($P$) for excited states.

In the CC($P$) computations, which are the first step of the CC($P$;$Q$) 
algorithm, we determine the cluster operator
\begin{equation}
T^{(P)} = \sum_{|\Phi_K\rangle \in \mathscr{H}^{(P)}} t_K E_K ,
\label{Toperator}
\end{equation}
associated with the CC($P$) ket state $|\Psi_{0}^{(P)}\rangle = e^{T^{(P)}} | \Phi\rangle$, and the
corresponding ground-state energy $E_{0}^{(P)} = \langle \Phi | \overline{H}^{(P)} | \Phi \rangle$,
where $\overline{H}^{(P)} = e^{-T^{(P)}} H e^{T^{(P)}}$. In the subsequent EOMCC($P$)
step, we diagonalize the similarity-transformed Hamiltonian $\overline{H}^{(P)}$ in the
$P$ space to obtain the excited-state energies $E_{\mu}^{(P)}$ and the corresponding linear
excitation operators
\begin{equation}
R_{\mu}^{(P)} = r_{\mu,0} {\bf 1} + \sum_{|\Phi_K\rangle  \in
\mathscr{H}^{(P)}} r_{\mu,K} E_K
\label{Roperator}
\end{equation}
defining the EOMCC($P$) ket states
$|\Psi_{\mu}^{(P)}\rangle = R_{\mu}^{(P)}e^{T^{(P)}} |\Phi\rangle$.
For both the ground ($\mu = 0$) and  excited ($\mu > 0$) states,
we also solve for the hole--particle deexcitation operators
\begin{equation}
L_{\mu}^{(P)} = \delta_{\mu,0} {\bf 1} +
\sum_{|\Phi_K\rangle \in \mathscr{H}^{(P)}} l_{\mu,K} (E_K)^{\dagger} ,
\label{Loperator}
\end{equation}
which define the companion CC($P$)/EOMCC($P$) bra wave functions
$\langle \tilde {\Psi}_{\mu}^{(P)} | = \langle \Phi | L_{\mu}^{(P)} e^{-T^{(P)}}$
satisfying the biorthonormality condition
$\langle\tilde{\Psi}_{\mu}^{(P)}|\Psi_{\nu}^{(P)}\rangle = \delta_{\mu,\nu}$.
Once the $T^{(P)}$, $R_{\mu}^{(P)}$, and $L_{\mu}^{(P)}$ operators and the
CC($P$) and EOMCC($P$) energies $E_{\mu}^{(P)}$ are determined, we calculate
the noniterative state-specific corrections
\begin{equation}
\delta_{\mu}(P;Q) = \sum_{|\Phi_K\rangle \in \mathscr{H}^{(Q)}} \ell_{\mu,K}(P)
\: \mathfrak{M}_{\mu,K}(P),
\label{mmcorrection}
\end{equation}
where $\mathfrak{M}_{0,K}(P) = \langle \Phi_K | \overline{H}^{(P)}|\Phi\rangle$ and
$\mathfrak{M}_{\mu,K}(P) = \langle \Phi_K | \overline{H}^{(P)} R_{\mu}^{(P)} |\Phi\rangle$
are the generalized moments of the CC($P$) ($\mu = 0$) \cite{leszcz,ren1}
and EOMCC($P$) ($\mu > 0$) \cite{kkppeom2} equations, which correspond to projections
of these equations on the $Q$-space determinants $|\Phi_{K}\rangle \in \mathscr {H}^{(Q)}$,
and the coefficients $\ell_{\mu,K}(P)$ multiplying moments $\mathfrak{M}_{\mu,K}(P)$
in Eq.\ (\ref{mmcorrection}) are defined as 
$\ell_{\mu,K}(P) = \langle \Phi| L_{\mu}^{(P)}\overline{H}^{(P)} |\Phi_K \rangle / D_{\mu,K}^{(P)}$, 
with $D_{\mu,K}^{(P)} = E_{\mu}^{(P)} - \langle \Phi_K | \overline{H}^{(P)} | \Phi_K \rangle$ 
representing the Epstein--Nesbet-like energy denominators. The final
CC($P$;$Q$) energies are calculated as
\begin{equation}
E_{\mu}^{(P + Q)} = E_{\mu}^{(P)} + \delta_{\mu}(P;Q) .
\label{ccpq_energy}
\end{equation}

A key advantage of the CC($P$;$Q$) framework is the flexibility it offers in
defining the $P$ and $Q$ spaces to accurately and efficiently capture the
many-electron correlation effects relevant to the electronic states of interest.
The present study exploits this freedom to examine the ground- and excited-state 
PESs of the water molecule along the O--H bond-breaking coordinate using three
different types of the CC($P$;$Q$) methodology aimed at accurately approximating the
high-level CCSDT/EOMCCSDT energetics.
In the first CC($P$;$Q$) approach examined in this work, the $P$ space is spanned by all singly
and doubly excited determinants, $|\Phi_{i}^{a}\rangle$ and $|\Phi_{ij}^{ab}\rangle$,
respectively, where $i, j, \, \ldots$ ($a, b, \, \ldots$) denote the occupied
(unoccupied) spinorbitals in $|\Phi\rangle$, and the $Q$ space consists of all
triply excited determinants $|\Phi_{ijk}^{abc} \rangle$. This conventional
variant of CC($P$;$Q$) represents the noniterative triples corrections to the
CCSD and EOMCCSD energies defining the ground-state CR-CC(2,3) method
\cite{crccl_jcp,crccl_cpl,crccl_molphys,crccl_jpc,jspp-chemphys2012} and its excited-state
CR-EOMCC(2,3) counterpart \cite{crccl_molphys,crccl_ijqc2,jspp-chemphys2012}, which were used
to determine the ground- and excited-state potentials of water in Ref.\ \cite{water}.
In the second CC($P$;$Q$) method, the $P$ and $Q$ spaces are
constructed with the help of active orbitals such that the $P$ space contains
all singly and doubly excited determinants and the subset of triply excited determinants
selected according to the formula $|\Phi_{ij{\bf K}}^{{\bf A}bc}\rangle$, where
${\bf K}$ (${\bf A}$) designate the occupied (unoccupied) spinorbitals around the
Fermi level belonging to the user-specified active spinorbital set, and the
$Q$ space is spanned by the remaining triply excited determinants. This choice of
$P$ and $Q$ spaces results in the CC(t;3) method of Refs.\
\cite{jspp-chemphys2012,jspp-jcp2012}, in which the ground-state CCSDt
\cite{semi0b,semi2,semi4,piecuch-qtp} and excited-state EOMCCSDt
\cite{eomccsdt1,eomccsdt2,piecuch-qtp} energies are
corrected for the remaining, largely dynamical, $T_{3}$ and $R_{\mu,3}$
correlation effects not captured by the CCSDt/EOMCCSDt calculations.
In the third CC($P$;$Q$) variant considered in this study, the $P$ space is spanned by all singly
and doubly excited determinants and the leading triply excited determinants
identified with the help of their $\ell_{\mu,K}(P)\:\mathfrak{M}_{\mu,K}(P)$
contributions to the $\delta_{\mu}(P;Q)$ corrections, Eq. (\ref{mmcorrection}),
using the adaptive CC($P$;$Q$) algorithm introduced in Ref.\ \cite{adaptiveccpq2023},
and the $Q$ space consists of the remaining triply excited determinants not included
in the $P$ space. Each of the above CC($P$;$Q$) approaches aims
at converging the CCSDT/EOMCCSDT energetics using the same general
recipe defined by Eqs. (\ref{Toperator}) -- (\ref{ccpq_energy}).
The only difference between them is in the method used to partition the manifold
of triple excitations between the iterative and noniterative steps of the
CC($P$;$Q$) calculations.
All three CC($P$;$Q$) approaches also offer major savings in the computational effort
compared to the full treatment of triples, which were, for example, discussed, along with
illustrative timings, in Refs.\ \cite{adaptiveccpq2023,ccpq-be2-jpca-2018}
(cf. Ref.\ \cite{stochastic-ccpq-molphys-2020} for additional relevant remarks).
It is, therefore, interesting to examine which of these
three CC($P$;$Q$) schemes is most effective in capturing
$T_{3}$ and $R_{\mu,3}$ correlations.
\section{Results and Discussion}
\label{sec3}
\subsection{Computational Details}
\label{sec3.1}
The nuclear geometries needed to construct the ground- and excited-state PESs of the water
molecule along the $\rm{H_{2}O} \rightarrow \rm{H} + \rm{OH}$ dissociation path,
obtained by considering 11 values of the O--H bond separation $R_{\rm{OH}}$
ranging from 1.3 to 4.4 bohr, with the remaining O--H bond length and H--O--H bond angle
optimized using the CCSD/cc-pVTZ method, were taken from Ref.\ \cite{nmapp3}.
As pointed out in Refs.\ \cite{water,nmapp3}, the resulting dissociation pathway on the
ground-state PES has several desirable characteristics. For example, the equilibrium
values of $R_{\rm{OH}}$ and H--O--H bond angle resulting from the CCSD/cc-pVTZ geometry
optimization, of 1.809 bohr and 103.9 degree, respectively, are in good agreement with
experiment and, as $R_{\rm{OH}}$ increases, the second
O--H bond length approaches its equilibrium value in the ground-state
OH radical and the angular potential flattens.
For each water structure corresponding to a particular value of $R_{\rm{OH}}$,
we carried out the CR-CC(2,3)/CR-EOMCC(2,3), active-orbital-based CC($P$;$Q$)
[i.e., CC(t;3)], adaptive CC($P$;$Q$), and parent CCSDT/EOMCCSDT calculations 
for the four lowest $\text{A}^{\prime}(C_{\rm s})$-symmetric singlet states, which include the
ground state \AP{X}{1} and \AP{n}{1} excited states with $n = 1\mbox{--}3$,
three lowest $\text{A}^{\prime}(C_{\rm s})$-symmetric triplet states 
(denoted as \AP{n}{3}, $n = 1\mbox{--}3$), two lowest
$\text{A}^{\prime\prime}(C_{\rm s})$-symmetric singlet states
(denoted as \APP{n}{1}, $n = 1$, 2), and
three lowest triplet states of the $\text{A}^{\prime\prime}(C_{\rm s})$ symmetry (denoted as
\APP{n}{3}, $n = 1\mbox{--}3$). To enable comparisons of our CCSDT/EOMCCSDT
data with the
full CI and MRCC energetics obtained in
Refs.\ \cite{nmapp3} (full CI and MRCC) and \cite{water} (full CI), all
calculations reported in this work were performed with the TZ basis set of
Ref.\ \cite{nmapp3}, used in Ref.\ \cite{water} as well. The restricted
Hartree--Fock (RHF) determinant, used in our CC/EOMCC calculations as a reference,
consisted of four $a^{\prime}(C_{\rm s})$-symmetric and one $a^{\prime\prime}(C_{\rm s})$-symmetric 
orbitals. The lowest-energy orbital correlating with 1s shell of oxygen was frozen in
post-RHF steps. To define the CC(t;3) and underlying CCSDt/EOMCCSDt calculations,
the three highest occupied and two lowest unoccupied orbitals in the RHF reference
correlating with 2p shell of oxygen and 1s shells of hydrogens
were treated as active. Two of the three active occupied and both active unoccupied
orbitals were of the $a^{\prime}(C_{\rm s})$ symmetry. The third active occupied
orbital was $a^{\prime\prime}(C_{\rm s})$-symmetric. The adaptive CC($P$;$Q$) calculations
employed the relaxed
algorithm, discussed in detail in Ref.\ \cite{adaptiveccpq2023}, in which we assumed
a 1\% growth rate in the numbers of triply excited determinants entering 
the underlying $P$ spaces. The percentages of triply excited determinants
characterizing our adaptive CC($P$), EOMCC($P$), and CC($P$;$Q$) calculations
were defined as fractions of the $S_z=0$ triples of the 
$\text{A}^{\prime}(C_{\rm s})$ ($^{1}\text{A}^{\prime}$ and $^{3}\text{A}^{\prime}$ states) 
or $\text{A}^{\prime\prime}(C_{\rm s})$ 
($^{1}\text{A}^{\prime\prime}$ and $^{3}\text{A}^{\prime\prime}$ states) 
symmetry identified with the adaptive CC($P$;$Q$) algorithm.
The CCSD, EOMCCSD, CR-CC(2,3), CR-EOMCC(2,3), CCSDt, EOMCCSDt, CC(t;3), CCSDT, and EOMCCSDT
computations were performed using
our in-house CC/EOMCC codes interfaced with the RHF and integral 
transformation routines in GAMESS  \cite{gamess2020}.
The adaptive CC($P$;$Q$)
calculations were executed using our recently developed CCpy package available on GitHub \cite{CCpy-GitHub}
that can efficiently handle the potentially spotty subsets of triply excited determinants
entering the underlying CC($P$) and EOMCC($P$) computations, which may not form continuous manifolds
labelled by occupied and unoccupied orbitals from the respective ranges of indices, to achieve
the desired speedups compared to CCSDT/EOMCCSDT (see Ref.\ \cite{adaptiveccpq2023} for further
information).

The results of our CC/EOMCC calculations are shown in Tables \ref{table1} -- \ref{table3}, 
Fig.\ \ref{figure1}, and the Supplementary Data document (see Appendix A).
Table \ref{table1} summarizes the mean unsigned error (MUE) and nonparallelity error (NPE)
values characterizing the CCSDT and EOMCCSDT potential cuts of water computed in this work
relative to their full CI counterparts obtained in Refs.\ \cite{nmapp3,water}.
The MUE and NPE values characterizing the CCSD/EOMCCSD, CR-CC(2,3)/CR-EOMCC(2,3),
adaptive CC($P$)/EOMCC($P$) and CC($P$;$Q$), CCSDt/EOMCCSDt, and CC(t;3) PESs
relative to their CCSDT/EOMCCSDT parents are reported in Tables \ref{table2} and
\ref{table3}, respectively. The ground- and excited-state PES cuts of water
corresponding to the $\mathrm{H_2O} \rightarrow \mathrm{H} + \mathrm{OH}$ dissociation
channels that correlate with the $X\;^{2}\Pi$ ground state and the lowest-energy
$^{2}\Sigma^{+}$ and $^{2}\Sigma^{-}$ states of $\mathrm{OH}$ obtained with
CCSD/EOMCCSD, CR-CC(2,3)/CR-EOMCC(2,3), adaptive CC($P$)/EOMCC($P$) and CC($P$;$Q$),
CCSDt/EOMCCSDt, CC(t;3), and CCSDT/EOMCCSDT are shown in Fig.\ \ref{figure1}.
The Supplementary Data document provides total electronic energies of all the
calculated states.
\subsection{Accuracy of the CCSDT and EOMCCSDT Approaches}
\label{sec3.2}
Given that all three variants of the CC($P$;$Q$) methodology tested in this work aim at 
converging or accurately approximating the CCSDT/EOMCCSDT energetics, it is important to assess
how well the CCSDT and EOMCCSDT approaches perform in describing the ground- and excited-state
PESs of water along the O--H bond-breaking coordinate relative to their exact, full CI, counterparts.
As shown in Table \ref{table1}, the CCSDT method provides a highly accurate description of the 
\AP{X}{1} state, with the MUE and NPE values relative to the ground-state full CI potential
being as small as 0.63 and 1.14 millihartree, respectively. EOMCCSDT is similarly accurate in describing
the excited-state potentials, with only three -- out of 11 examined in this study -- characterized
by the MUE and NPE values exceeding 1 and 3  millihartree, respectively, but even in this case,
which includes the \AP{2}{1}, \AP{3}{1}, and \AP{3}{3}
states, the MUE and NPE values, of 1.28 and 3.19 millihartree for the \AP{2}{1} PES,
1.53 and 4.48 millihartree for the \AP{3}{1} PES, and 1.41 and 5.50 millihartree
for the \AP{3}{3} PES, remain rather small, especially when we realize that these
three states are located hundreds of millihartrees above the ground state (cf. Supplementary
Data). Most importantly, the EOMCCSDT approach accurately approximates the full CI water
potentials in the highly stretched $R_{\rm{OH}} \geq 2.4$ bohr region, where all excited 
states of interest in this study acquire a substantial MR character that EOMCCSD cannot capture.
When compared with the MRCC results reported in Ref. \cite{nmapp3}, we observe that the
SR CCSDT and EOMCCSDT methods can also rival, or even outperform, the state-of-the-art MR treatments.
For example, for the \AP{1}{3}, \AP{1}{1}, \APP{2}{3},
\APP{2}{1}, \AP{3}{1}, and \AP{3}{3} states,
the EOMCCSDT NPE values relative to full CI are lower -- sometimes substantially --
than those resulting from the $n$R-GMS-SU-CCSD and $(N,M)$-CCSD calculations.
For the \AP{X}{1}, \APP{1}{1}, \APP{1}{3}, \AP{2}{3}, and \AP{2}{1} potentials, 
the CCSDT and EOMCCSDT
NPE values relative to full CI are within $\sim$1 millihartree from the best MRCC
results reported in Ref. \cite{nmapp3}. It is clear from these comparisons that
we can treat the ground- and excited-state potentials of water along the O--H
bond-breaking coordinate obtained with CCSDT and EOMCCSDT as highly accurate benchmarks
for evaluating performance of the different CC($P$;$Q$) approaches explored in this work.
\subsection{Performance of Different CC(\protect\mbox{$P$;$Q$}) Approaches 
Relative to CCSDT and EOMCCSDT}
\label{sec3.3}
We begin by discussing the first variant of the CC($P$;$Q$) methodology of interest
in the present study corresponding to the CR-CC(2,3) and CR-EOMCC(2,3) triples corrections
to CCSD and EOMCCSD (also examined in Ref.\ \cite{water}).
In the vicinity of the equilibrium geometry on the ground-state PES, all
CC/EOMCC approaches, including CR-CC(2,3)/CR-EOMCC(2,3), and even CCSD/EOMCCSD,
perform well. Indeed, the errors relative to CCSDT in the \AP{X}{1} potential computed with CCSD
in the $R_{\rm OH} = 1.3\mbox{--}2.0$ bohr region range between 2.771 and 3.562 millihartree.
CR-CC(2,3) reduces them to $\sim$0.2--0.3 millihartree. For the
excited-state potentials in the $R_{\rm OH} = 1.3\mbox{--}2.0$ bohr region, the largest error
obtained with EOMCCSD relative to EOMCCSDT, which occurs when the high-lying \AP{3}{1} state
is considered, is 3.392 millihartree. The largest error characterizing CR-EOMCC(2,3) in the same
region, encountered when the \APP{3}{3} PES is examined, is 1.660 millihartree. For nearly
all excited states considered in this work, the differences between the CR-EOMCC(2,3) and EOMCCSDT
potentials in the $R_{\rm OH} = 1.3\mbox{--}2.0$ bohr region are about 1 millihartree, making
them virtually identical around the equilibrium geometry.
The CR-EOMCC(2,3) PESs are also more parallel to their EOMCCSDT counterparts
than the corresponding EOMCCSD surfaces.

\begin{figure}
\centering
\includegraphics[scale=0.68]{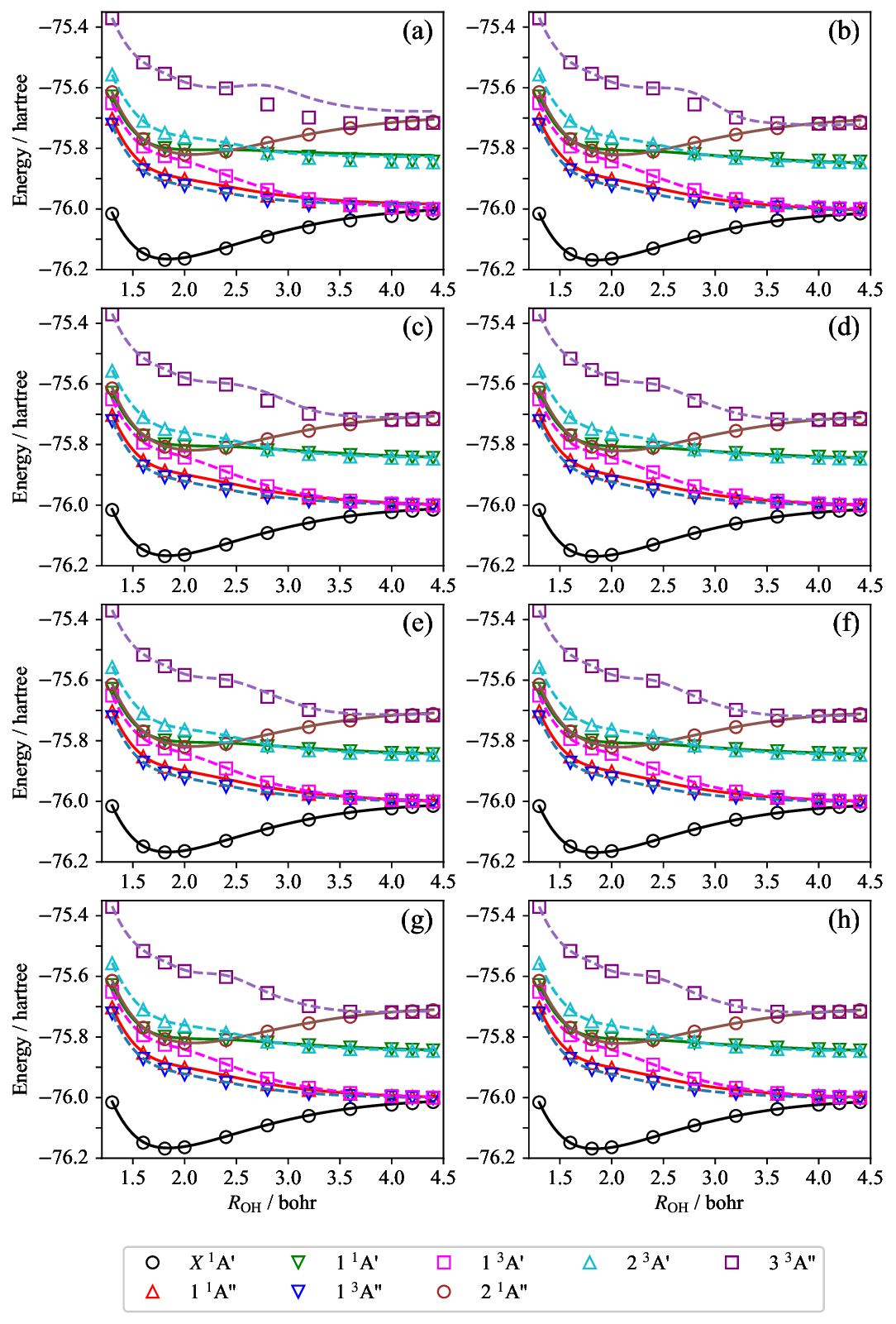}
\input{figure1-caption}
\label{figure1}
\end{figure}

The situation changes when we move toward larger $R_{\rm OH}$ values, where all electronic states
of water considered in this study develop a strong MR character, resulting in failures of the
CCSD and EOMCCSD methods, especially when the
\AP{X}{1}, \APP{1}{1}, \APP{1}{3}, \AP{1}{1}, \AP{2}{3},
\APP{2}{3}, \AP{2}{1}, \APP{3}{3}, \AP{3}{1}, and \AP{3}{3}
potentials
are examined. These failures become particularly dramatic for
the \APP{2}{3}, \AP{2}{1}, \APP{3}{3}, and \AP{3}{1} states, where
errors relative to EOMCCSDT resulting from the EOMCCSD calculations in the $R_{\rm OH} = 2.8\mbox{--}4.4$
bohr region become as large as 28.470, 32.540, 65.442, and 32.035 millihartree, respectively, although no
state considered in our computations is accurately described when
$T_{3}$ and $R_{\mu,3}$ correlations are neglected and $R_{\rm OH} > 2.4$ bohr. The
CR-CC(2,3) and CR-EOMCC(2,3) triples corrections reduce the excessive errors characterizing the
\AP{X}{1}, \APP{1}{1}, \APP{1}{3}, \AP{1}{1}, \AP{2}{3},
\APP{2}{3}, \AP{2}{1}, \APP{3}{3}, \AP{3}{1}, and \AP{3}{3}
potentials obtained with CCSD/EOMCCSD at larger values of $R_{\rm OH}$ and the associated
MUEs and NPEs shown in Tables \ref{table2} and \ref{table3}, which are 6.154--28.262 and 8.453--64.663
millihartree, respectively, in the CCSD/EOMCCSD case and 0.543--6.075 and 0.692--42.170 millihartree
when the CR-CC(2,3)/CR-EOMCC(2,3) data are examined, but several problems remain, especially when
dealing with the \APP{2}{3} and \APP{3}{3} states.
In the former case, the CR-EOMCC(2,3) approach produces large, 11.891--13.976 millihartree,
errors relative to EOMCCSDT in the $R_{\rm OH} = 4.0\mbox{--}4.4$ bohr region, resulting in
the qualitatively incorrect asymptotic behavior of the CR-EOMCC(2,3) \APP{2}{3} potential
and the NPE of 13.043 millihartree.
The CR-EOMCC(2,3) PES for the \APP{3}{3} state presents an even more
distressing situation, with a massive, 37.018 millihartree, error relative to 
EOMCCSDT at $R_{\rm OH} = 2.8$ bohr associated with a bump in the \APP{3}{3} potential
seen in Fig.\ \ref{figure1}(b) and the even larger NPE value of 42.170 millihartree.
These failures of CR-EOMCC(2,3) are consistent with our earlier observations
\cite{jspp-chemphys2012,jspp-jcp2012,stochastic-ccpq-prl-2017,stochastic-ccpq-molphys-2020,%
cipsi-ccpq-2021,adaptiveccpq2023}
that none of the triples corrections to CCSD/EOMCCSD can provide accurate results
when $T_{3}$ and $R_{\mu,3}$ correlations become substantial and
strongly coupled to their lower-rank 
counterparts, as is the case when examining the \APP{2}{3} and \APP{3}{3} potentials
at larger $R_{\rm OH}$ values.

The above challenges encountered in the CR-EOMCC(2,3) calculations can be addressed by
turning to the CC(t;3) method. As explained in Sec.\ \ref{sec2}, CC(t;3) is a
CC($P$;$Q$) approach in which energies obtained in the CCSDt/EOMCCSDt calculations
that incorporate the leading triply excited determinants identified with the help of
active orbitals in the $P$ spaces used in the iterative CC($P$)/EOMCC($P$) steps
are corrected for the missing, mostly dynamical, $T_{3}$ and $R_{\mu,3}$ correlations
using Eq. (\ref{mmcorrection}). The numerical results in Tables \ref{table2} and
\ref{table3}, with further details provided by Tables S2--S13 in Supplementary Data,
and the potential curves shown in Fig.\ \ref{figure1}(h) demonstrate that the CC(t;3)
method readily addresses the shortcomings of the CR-EOMCC(2,3) approach and its
ground-state CR-CC(2,3) counterpart. This becomes particularly clear when examining
the \APP{2}{3} and \APP{3}{3} potentials that are poorly
described by CR-EOMCC(2,3). For example, the CC(t;3) calculations reduce the
large, 11.891--13.976 millihartree, errors relative to EOMCCSDT obtained with
CR-EOMCC(2,3) for the \APP{2}{3} PES in the $R_{\rm OH} = 4.0\mbox{--}4.4$ bohr region
to 0.481--0.625 millihartree. The massive error of 37.018 millihartree produced
by CR-EOMCC(2,3) for the \APP{3}{3} state at $R_{\rm OH} = 2.8$ bohr is reduced in
the CC(t;3) calculations by more than two orders of magnitude, to 0.133 millihartree,
eliminating the unphysical bump in the CR-EOMCC(2,3) potential for this state
in the $R_{\rm OH} = 2.4\mbox{--}3.2$ bohr region altogether. The major improvements
offered by CC(t;3) in describing the \APP{2}{3} and \APP{3}{3} PESs are
also reflected in the corresponding NPE values relative to EOMCCSDT, which decrease
from 13.043 and 42.170 millihartree obtained with CR-EOMCC(2,3) to the minuscule
0.255 and 0.608 millihartree, respectively. In general, the CC($P$;$Q$)-based
CC(t;3) method provides a highly accurate description of the ground and excited
states of water along the O--H bond-breaking coordinate, greatly improving the
CR-CC(2,3)/CR-EOMCC(2,3) energetics and closely matching the nearly exact
CCSDT/EOMCCSDT potentials
at small fractions of the computational costs associated with CCSDT/EOMCCSDT.
This is manifested by the
tiny MUE and NPE values relative to CCSDT/EOMCCSDT characterizing the
CC(t;3) calculations for the 12 electronic states of water reported in this work,
which range from 0.166 to 0.951 millihartree for MUEs and 0.110 to
0.608 millihartree for NPEs (see Tables \ref{table2} and \ref{table3}).
By capturing the missing $T_{3}$ and $R_{\mu,3}$ correlations, the
CC(t;3) corrections are also very effective in improving the underlying
CCSDt/EOMCCSDt potentials, especially in reducing the MUE values relative to
CCSDT/EOMCCSDT characterizing the CCSDt and EOMCCSDt calculations, from
1.477--2.420 millihartree in CCSDt/EOMCCSDt to fractions of a millihartree
in CC(t;3).

We conclude
by commenting on the results of our
adaptive CC($P$;$Q$) calculations,
which are based on the same basic principles as those employed in the CC(t;3)
considerations, but
do not rely on
active orbitals to obtain accurate results, allowing us to converge the high-level
CCSDT and EOMCCSDT energetics in an entirely black-box
fashion.
As shown in Tables \ref{table2} and \ref{table3}, panels (d) and (f) of Fig.\ \ref{figure1},
and Tables S2--S13 in Supplementary Data, the adaptive CC($P$;$Q$) approach is
remarkably effective in accurately approximating the CCSDT/EOMCCSDT
water potentials.
Already with a tiny 1\% of triply excited determinants
in the underlying $P$ spaces, the adaptive CC($P$;$Q$) calculations offer major
improvements in the CR-CC(2,3)/CR-EOMCC(2,3) data. They reduce the MUE and NPE
values of 0.543 and 0.692 millihartree, respectively, relative to CCSDT, obtained
with CR-CC(2,3) for the ground-state potential, to less than 0.3 millihartree.
The improvements in the description of the 11 excited states of water considered
in this work offered by the adaptive CC($P$;$Q$) approach using the leading 1\% of
triply excited determinants in the underlying $P$ spaces, which for the sake of brevity
of this discussion will be abbreviated as CC($P$;$Q$)[\%T = 1], are similarly impressive,
especially when we realize that in the case of the $C_{\rm s}$-symmetric water structures
and the TZ basis set used in our calculations, 1\% of triply excited determinants
amounts to only about 300 $T_{3}$ and $R_{\mu,3}$ amplitudes,
as opposed to 31,832 $\text{A}^{\prime}(C_{\rm s})$-symmetric and 32,232
$\text{A}^{\prime\prime}(C_{\rm s})$-symmetric $S_{z} = 0$ triples used by full
CCSDT/EOMCCSDT. These improvements can be best seen when comparing the
\APP{2}{3} and \APP{3}{3} potentials obtained with CR-EOMCC(2,3) with their
CC($P$;$Q$)[\%T = 1] counterparts. In the case of the \APP{2}{3} PES, the
MUE and NPE values of 5.033 and 13.043 millihartree relative to EOMCCSDT
resulting from the CR-EOMCC(2,3) computations are reduced to 1.177 and 1.081
millihartree, respectively, when the CC($P$;$Q$)[\%T = 1] approach is employed,
helping to alleviate the inaccurate behavior of CR-EOMCC(2,3) in the asymptotic part
of the \APP{2}{3} potential. The poor description of the \APP{3}{3} PES by
the CR-EOMCC(2,3) method becomes much more reasonable in the CC($P$;$Q$)[\%T = 1]
calculations as well. The 37.018 millihartree error relative to EOMCCSDT
obtained with CR-EOMCC(2,3) at $R_{\rm OH} = 2.8$ bohr reduces to 6.740 millihartree
when the adaptive CC($P$;$Q$)[\%T = 1] approach is employed. The MUE and NPE values
characterizing the CR-EOMCC(2,3) \APP{3}{3} potential, of 6.075 and 42.170 millihartree,
respectively, decrease in the CC($P$;$Q$)[\%T = 1] calculations to 1.949 and 5.933 millihartree
and, as shown in Fig. \ref{figure1} (d), the unphysical bump in the
CR-EOMCC(2,3) PES for this state in the $R_{\rm OH} = 2.4\mbox{--}3.2$ bohr region
disappears. The generally small MUE and NPE values characterizing the CC($P$;$Q$)[\%T = 1]
calculations for the 12 PESs of water considered in this study, relative to their
CCSDT/EOMCCSDT parents, which range from
0.275 to 1.949 millihartree for MUEs and 0.216 to 5.933 millihartree for NPEs,
are certainly encouraging. As shown in Tables \ref{table2} and \ref{table3},
Fig.\ \ref{figure1}(f), and Tables S2--S13 in Supplementary Data, the situation
gets even better when the fraction of triply excited determinants included in the
$P$ spaces defining the adaptive CC($P$;$Q$) calculations increases to 2\%.
The MUEs and NPEs relative to CCSDT/EOMCCSDT characterizing the resulting
CC($P$;$Q$)[\%T = 2] computations for the 12 potential cuts of water examined
in this work reduce to 0.197--0.993 and 0.133--1.572 millihartree, respectively. In
the case of the \APP{2}{3} and \APP{3}{3} potentials that cause major troubles to the
CR-EOMCC(2,3) method, the already small MUEs obtained in the CC($P$;$Q$)[\%T = 1]
calculations, of 1.177 and 1.949 millihartree, decrease to 0.737 and 0.993 millihartree,
respectively, when the fraction of triply excited determinants incorporated in the
underlying $P$ spaces grows from 1\% to 2\%. The corresponding NPE values decrease
from 1.081 and 5.933 millihartree in the CC($P$;$Q$)[\%T = 1] case to 0.237 and 1.572
millihartree, when the CC($P$;$Q$)[\%T = 2] approach is employed. For some electronic
potentials of water, the NPEs characterizing the adaptive CC($P$;$Q$)[\%T = 2] computations
are slightly larger than those obtained with the CC(t;3) approach, but the overall
performance of the adaptive CC($P$;$Q$)[\%T = 2] and CC(t;3) methods is very similar.
This is promising for the future applications of the adaptive CC($P$;$Q$)
framework since typical CC(t;3) computations, in addition to requiring the user to
select active orbitals, employ considerably larger
fractions of triply excited determinants in the iterative steps of the underlying
CC($P$;$Q$) algorithm than their black-box adaptive CC($P$;$Q$) counterparts.
For example, the CC(t;3) calculations reported in this work used about 38\% of all
triples in the iterative CCSDt/EOMCCSDt steps, as opposed to only 2\% used in the
adaptive CC($P$;$Q$) computations producing virtually identical results.
\section{Summary}
\label{sec4}
We extended the active-orbital-based and adaptive CC($P$;$Q$) methodologies
of Refs.\ \cite{jspp-chemphys2012,jspp-jcp2012,adaptiveccpq2023} to excited
electronic states and used the resulting approaches aimed at converging the
CCSDT and EOMCCSDT energetics to determine the ground- and excited-state
potentials of the water molecule along the O--H bond-breaking coordinate,
for which CCSDT and EOMCCSDT accurately approximate the full CI data. We
demonstrated that the active-orbital-based CC($P$;$Q$) method, abbreviated
as CC(t;3), and its black-box adaptive CC($P$;$Q$) counterpart are similarly
effective in accurately approximating the parent CCSDT/EOMCCSDT energetics at
small fractions of the computational costs,
improving the results obtained with the CR-CC(2,3) and CR-EOMCC(2,3) triples
corrections to CCSD/EOMCCSD, but the adaptive CC($P$;$Q$) approach
produced potentials of the CCSDT/EOMCCSDT
quality with much smaller fractions of triply excited determinants in the
underlying $P$ spaces than those used by CC(t;3).
In our future work, we will investigate if the excellent performance of the
active-orbital-based and adaptive CC($P$;$Q$) methods demonstrated in this
study holds for other molecular systems and larger basis sets, for which full CI
and CCSDT/EOMCCSDT calculations, used in this work to assess accuracies
of the various CC($P$;$Q$) approaches, are no longer possible.
\section*{CRediT authorship contribution statement}
\textbf{Karthik Gururangan}: Methodology, Software, Data curation, Formal 
analysis, Validation, Writing - original draft.
\textbf{Jun Shen}: Methodology, Software, Data curation, Formal 
analysis, Validation, Writing - original draft.
\textbf{Piotr Piecuch}: Conceptualization, Methodology, Formal 
analysis, Investigation, Funding acquisition, Project administration, 
Resources, Supervision, Validation, Writing - reviewing and editing.
\section*{Declaration of competing interest}
The authors declare that they have no known competing financial
interests or personal relationships that could have appeared to influence
the work reported in this paper.
\section*{Data availability}
The data that support the findings of this study are available
within the article and the Supplementary Data.
\section*{Acknowledgments}
This work has been supported by the Chemical Sciences,
Geosciences and Biosciences Division, 
Office of Basic Energy Sciences, 
Office of Science, 
U.S. Department of Energy 
(Grant No. DE-FG02-01ER15228 to P.P).
\section*{Appendix A. Supplementary data}
Supplementary data to this article can be found online at 

\bibliographystyle{elsarticle-num}
\biboptions{sort,compress}
\bibliography{refs}

\newpage
\clearpage
\pagebreak
\input{table1}

\newpage
\clearpage
\pagebreak
\input{table2}

\newpage
\clearpage
\pagebreak
\input{table3}

%
%
\end{document}

%% file: table1.tex
\begin{table*}[h]
\centering
\footnotesize
\begin{threeparttable}
\caption{
The MUE and NPE values, in millihartree, relative to full CI 
characterizing the ground-state CCSDT and excited-state EOMCCSDT
potentials of the water molecule, as described by the TZ
basis set of Ref.\ \cite{nmapp3}, along the O--H bond-breaking coordinate
corresponding to the $\rm{H_{2}O} \rightarrow \rm{H} + \rm{OH}$ dissociation.
}
\label{table1}
\begin{tabular}{lcccccccccccc}
\hline\hline
 & \AP{X}{1} & \APP{1}{1} & \AP{1}{3} & \APP{1}{3} & \AP{1}{1} & \AP{2}{3} & \APP{2}{3} & \AP{2}{1} & \APP{2}{1} & \APP{3}{3} & \AP{3}{1} & \AP{3}{3} \\
\hline
MUE & 0.63 & 0.78 & 0.28 & 0.90 & 0.92 & 0.93 & 0.43 & 1.28 & 0.59 & 0.63 & 1.53 & 1.41\\
NPE & 1.14 & 1.85 & 0.83 & 2.14 & 2.13 & 2.33 & 1.35 & 3.19 & 0.57 & 2.67 & 4.48 & 5.50\\
\hline\hline
\end{tabular}
\end{threeparttable}
\end{table*}

%% file: table2.tex
\begin{table*}[h]
\centering
\footnotesize
\begin{threeparttable}
\caption{
The MUE values, in millihartree, relative to CCSDT/EOMCCSDT
characterizing the ground- and excited-state potential
cuts of the water molecule, as described by the TZ basis set of Ref.\ \cite{nmapp3},
along the O--H bond-breaking coordinate
corresponding to the $\rm{H_{2}O} \rightarrow \rm{H} + \rm{OH}$ dissociation
obtained with the different CC($P$)/EOMCC($P$) and CC($P$;$Q$) approaches examined in
the present work.
}
\label{table2}
\begin{tabular}{l c c c c @{\extracolsep{0.16in}} c @{\extracolsep{0.12in}} c @{\extracolsep{0.16in}} c @{\extracolsep{0.12in}} c}
\hline\hline
& & & & & \multicolumn{2}{c}{\%T = 1\tnote{e}} & \multicolumn{2}{c}{\%T = 2\tnote{f}} \\
\cline{6-7}
\cline{8-9}
State & CCSD\tnote{a} & CR(2,3)\tnote{b} & CCSDt\tnote{c} & CC(t;3)\tnote{d} & CC($P$) & CC($P$;$Q$) & CC($P$) & CC($P$;$Q$) \\
\hline
\AP{X}{1}   & 6.154  & 0.543 & 2.047 & 0.166 & 2.176 & 0.275 & 1.616 & 0.197 \\
\APP{1}{1}  & 7.666  & 1.517 & 1.683 & 0.599 & 3.159 & 0.755 & 2.271 & 0.622 \\
\AP{1}{3}   & 2.819  & 1.137 & 1.741 & 0.951 & 2.061 & 0.793 & 1.532 & 0.616 \\
\APP{1}{3}  & 7.693  & 1.080 & 1.658 & 0.627 & 3.237 & 0.854 & 2.339 & 0.665 \\
\AP{1}{1}   & 10.103 & 1.544 & 1.602 & 0.589 & 3.578 & 0.590 & 2.420 & 0.494 \\
\AP{2}{3}   & 8.846  & 1.225 & 1.682 & 0.646 & 3.379 & 0.665 & 2.322 & 0.530 \\ 
\APP{2}{3}  & 9.258  & 5.033 & 1.565 & 0.590 & 5.384 & 1.177 & 3.532 & 0.737 \\
\AP{2}{1}   & 14.460 & 2.988 & 2.276 & 0.718 & 3.230 & 0.782 & 2.359 & 0.618 \\
\APP{2}{1}  & 2.337  & 1.750 & 1.477 & 0.553 & 3.461 & 0.870 & 2.535 & 0.579 \\
\APP{3}{3}  & 28.262 & 6.075 & 2.420 & 0.542 & 8.111 & 1.949 & 4.973 & 0.993 \\
\AP{3}{1}   & 13.547 & 2.410 & 1.688 & 0.804 & 5.151 & 1.152 & 3.120 & 0.643 \\
\AP{3}{3}   & 8.305  & 2.481 & 1.656 & 0.854 & 3.794 & 1.048 & 2.671 & 0.676 \\
\hline\hline
\end{tabular}
\begin{tablenotes}
\footnotesize
\item[a]{
CCSD for the ground state and EOMCCSD for excited states.
}
\item[b]{
CR-CC(2,3) for the ground state and CR-EOMCC(2,3) for excited states.
}
\item[c]{
CCSDt/EOMCCSDt calculations using the active space consisting of
the three highest occupied and two lowest unoccupied RHF orbitals.
}
\item[d]{
CC(t;3) calculations using the active space consisting of
the three highest occupied and two lowest unoccupied RHF orbitals.
}
\item[e]{
CC($P$)/EOMCC($P$) and CC($P$;$Q$) calculations using $P$ spaces consisting of all
singly and doubly excited determinants and 1\% of triply excited determinants
identified by the adaptive CC($P$;$Q$) algorithm.
}
\item[f]{
CC($P$)/EOMCC($P$) and CC($P$;$Q$) calculations using $P$ spaces consisting of all
singly and doubly excited determinants and 2\% of triply excited determinants
identified by the adaptive CC($P$;$Q$) algorithm.
}
\end{tablenotes}
\end{threeparttable}
\end{table*}

%% file: table3.tex
\begin{table*}[h]
\centering
\footnotesize
\begin{threeparttable}
\caption{
The NPE values, in millihartree, relative to CCSDT/EOMCCSDT
characterizing the ground- and excited-state potential
cuts of the water molecule, as described by the TZ basis set of Ref.\ \cite{nmapp3},
along the O--H bond-breaking coordinate
corresponding to the $\rm{H_{2}O} \rightarrow \rm{H} + \rm{OH}$ dissociation
obtained with the different CC($P$)/EOMCC($P$) and CC($P$;$Q$) approaches examined in
the present work.
}
\label{table3}
\begin{tabular}{l c c c c @{\extracolsep{0.16in}} c @{\extracolsep{0.12in}} c @{\extracolsep{0.16in}} c @{\extracolsep{0.12in}} c}
\hline\hline
& & & & & \multicolumn{2}{c}{\%T = 1\tnote{e}} & \multicolumn{2}{c}{\%T = 2\tnote{f}} \\
\cline{6-7}
\cline{8-9}
State & CCSD\tnote{a} & CR(2,3)\tnote{b} & CCSDt\tnote{c} & CC(t;3)\tnote{d} & CC($P$) & CC($P$;$Q$) & CC($P$) & CC($P$;$Q$) \\
\hline
\AP{X}{1}    & 8.453  & 0.692 & 0.687 & 0.110 & 1.072 & 0.216 & 0.801 & 0.159 \\
\APP{1}{1}   & 16.093 & 4.592 & 0.446 & 0.275 & 1.569 & 0.433 & 1.304 & 0.198 \\
\AP{1}{3}    & 5.004  & 0.442 & 0.452 & 0.492 & 0.801 & 0.593 & 1.005 & 0.281 \\
\APP{1}{3}   & 17.445 & 3.303 & 0.447 & 0.240 & 2.070 & 0.430 & 1.360 & 0.133 \\
\AP{1}{1}    & 21.043 & 4.637 & 0.387 & 0.244 & 2.692 & 0.313 & 1.380 & 0.168 \\
\AP{2}{3}    & 20.127 & 3.712 & 0.826 & 0.257 & 2.264 & 0.246 & 1.233 & 0.218 \\
\APP{2}{3}   & 29.017 & 13.043& 0.572 & 0.255 & 6.852 & 1.081 & 2.761 & 0.237 \\
\AP{2}{1}    & 33.260 & 8.330 & 2.231 & 0.271 & 2.216 & 0.761 & 1.434 & 0.508 \\
\APP{2}{1}   & 7.971  & 3.983 & 0.601 & 0.263 & 3.129 & 0.611 & 0.837 & 0.636 \\
\APP{3}{3}   & 64.663 & 42.170& 3.601 & 0.608 & 23.376& 5.933 & 10.679& 1.572 \\
\AP{3}{1}    & 30.846 & 4.757 & 0.463 & 0.603 & 6.019 & 2.207 & 2.862 & 0.993 \\
\AP{3}{3}    & 21.562 & 5.670 & 0.712 & 0.392 & 3.149 & 1.574 & 1.413 & 0.582 \\ 
\hline\hline
\end{tabular}
\begin{tablenotes}
\footnotesize
\item[a]{
CCSD for the ground state and EOMCCSD for excited states.
}
\item[b]{
CR-CC(2,3) for the ground state and CR-EOMCC(2,3) for excited states.
}
\item[c]{
CCSDt/EOMCCSDt calculations using the active space consisting of
the three highest occupied and two lowest unoccupied RHF orbitals.
}
\item[d]{
CC(t;3) calculations using the active space consisting of
the three highest occupied and two lowest unoccupied RHF orbitals.
}
\item[e]{
CC($P$)/EOMCC($P$) and CC($P$;$Q$) calculations using $P$ spaces consisting of all
singly and doubly excited determinants and 1\% of triply excited determinants
identified by the adaptive CC($P$;$Q$) algorithm.
}
\item[f]{
CC($P$)/EOMCC($P$) and CC($P$;$Q$) calculations using $P$ spaces consisting of all
singly and doubly excited determinants and 2\% of triply excited determinants
identified by the adaptive CC($P$;$Q$) algorithm.
}
\end{tablenotes}
\end{threeparttable}
\end{table*}

%% file: figure1-caption.tex
\caption{
A comparison of the potential cuts of water, plotted as functions of $R_{\rm OH}$ and
corresponding to the $\mathrm{H_2O} \rightarrow \mathrm{H} + \mathrm{OH}$ dissociation 
channels that correlate with the $X\;^{2}\Pi$ ground state and the lowest-energy
$^{2}\Sigma^{+}$ and $^{2}\Sigma^{-}$ states of $\mathrm{OH}$, obtained with
(a) CCSD/EOMCCSD,
(b) CR-CC(2,3)/CR-EOMCC(2,3),
(c) adaptive CC($P$)/EOMCC($P$) using 1\% of triply excited determinants in the $P$ spaces,
(d) adaptive CC($P$;$Q$) using 1\% of triply excited determinants in the $P$ spaces,
(e) adaptive CC($P$)/EOMCC($P$) using 2\% of triply excited determinants in the $P$ spaces,
(f) adaptive CC($P$;$Q$) using 2\% of triply excited determinants in the $P$ spaces,
(g) CCSDt/EOMCCSDt, and
(h) CC(t;3)
with their full CCSDT/EOMCCSDT counterparts.
The splined CCSD/EOMCCSD, CR-CC(2,3)/CR-EOMCC(2,3),
adaptive CC($P$)/EOMCC($P$), adaptive CC($P$;$Q$), CCSDt/EOMCCSDt, and CC(t;3)
data are represented by the solid and dashed lines, whereas the open circles,
squares, triangles, and inverted triangles correspond to the parent
CCSDT/EOMCCSDT energetics.
}